\documentclass[a4paper]{jpconf}
\usepackage{graphicx}

\newcommand{\GeV}      {~\mathrm{GeV}}

\newcommand{\beqn}{\begin{eqnarray}}
\newcommand{\eeqn}{\end{eqnarray}}
\newcommand{\be}{\begin{equation}}
\newcommand{\ee}{\end{equation}}

\newcommand{\mathsym}[1]{{}}

\def \n34{\tilde{\chi}^{0}_{3,4}}

\def\met100{\slashed{E}_T\geq 100 \GeV}
\newcommand{\st}{Stueckelberg~}

\newcommand{\gappeq}{\mathrel{\rlap {\raise.5ex\hbox{$>$}}
{\lower.5ex\hbox{$\sim$}}}}
\newcommand{\lappeq}{\mathrel{\rlap{\raise.5ex\hbox{$<$}}
{\lower.5ex\hbox{$\sim$}}}}

\def\met{\slashed{E}_{T}}

\def\e{\end{document}}

\begin{document}
\title{SUGRA Grand Unification, LHC
and Dark Matter}
\author{Pran Nath }
\address{Department of Physics, Northeastern University, Boston, MA 02115, USA}
\ead{nath@neu.edu}

\begin{abstract}
A brief review is given of recent developments related  to the Higgs signal and its implications for
supersymmetry in the  supergravity grand unification framework.  The Higgs data indicates  
that the allowed parameter space largely lies on Focal Curves and Focal Surfaces of the Hyperbolic 
Branch of radiative breaking of the electroweak symmetry where TeV size scalars naturally arise.
The high mass of the Higgs leads to a more precise prediction for the allowed range of the 
spin independent neutralino-proton cross-section which is encouraging for the detection of dark
matter in future experiments with  greater sensitivity. Also discussed is the status of grand unification
and a natural solution to breaking the GUT group at one scale and  resolving the doublet-triplet
problem. It is shown that the cosmic coincidence  can be compatible within a supersymmetric 
framework in a muticomponent dark matter picture with one component charged under $B-L$ 
while the  other component is the conventional supersymmetric dark matter candidate, the neutralino.

\end{abstract}

\section{Introduction}
In December 2011 the ATLAS ~\cite{ATLAShiggs}  and CMS ~\cite{CMShiggs} Collaborations reported the result of  their updated search for the Higgs boson ~\cite{HiggsBoson}
which showed indications of a signal around the $3\sigma$ level. 
 These results lead to 
considerable theoretical activity to extract the implications of the signal ~\cite{Akula:2011aa,higgs_7tev1,higgs_7tev2}. 
More recently the analyses of  ~\cite{ATLAShiggs,CMShiggs} have received further support from
the Tevatron analyses~\cite{tevatron} and from the combined 7~TeV and 8~TeV CMS and ATLAS data~\cite{July4}.
Thus the  CMS Collaboration indicates a  signal for a boson with mass  of 
$125.3\pm 0.6  ~{\rm GeV}$ at $4.9\sigma$  and the ATLAS Collaboration indicates a signal for a boson
with a mass $\sim 126.5 ~{\rm GeV}$ at  $5.0\sigma$ in the combined   7~TeV and 8~TeV analysis~\cite{July4}
which  confirm the previous indication of the boson signal by ATLAS ~\cite{ATLAShiggs}  and CMS ~\cite{CMShiggs}.
The properties of the boson still need to be fully established including its spin. Assuming the particle discovered
is a spin 0 CP even boson, it is pertinent to ask if it is indeed the Higgs particle which enters in  spontaneous breaking  
of the electroweak symmetry and  gives mass to gauge bosons 
 and to quarks and leptons\cite{HiggsBoson}. Further, if indeed the  discovered particle is the Higgs boson, 
 it is of interest to determine its properties and any 
  possible deviations from the standard model predictions  which would be indicators of new physics beyond
  the standard model. 
Definitive answers to these questions will have to wait for much more data. However, partial answers already exist in the data 
 from the Tevatron and from the CMS/ATLAS experiments. 
 Thus  there is evidence already that the observed particle emulates the most obvious property of the Higgs boson, 
 i.e., that its coupling to heavier fermions is larger than its coupling to  lighter fermions.
 Thus, if the Higgs boson coupled with equal strength to fermions one would have seen an equal 
 abundance of $\mu^+\mu^-$ as of $b\bar b$. However, the Tevatron data suggests otherwise. While there is 
 evidence  of $b\bar b$ decay of the Higgs there is no evidence of Higgs decaying
 into $\mu^+\mu^-$~\cite{tevatron}.\\.  
 
\section{Higgs and SUSY}
An experimental determination of the Higgs couplings are essential in testing the standard model
and to find any deviations from the standard model prediction which would be an indication of new physics.
The relevant couplings to test are the Higgs boson  couplings to femion-anti-fermion pairs and
to dibosons, i.e., $ hf\bar f, ~hWW, ~hZZ, ~h \gamma\gamma, ~h  Z\gamma.$
Thus  at the LHC with $\sqrt{s}=14$ TeV, and ${\cal L} =300$ fb$^{-1}$  up to 
$10\%$ accuracy for couplings of $h^0$  with dibosons $WW, ZZ$ and up to $\sim 20\%$ for couplings to b, $\tau$
appear possible~\cite{Peskin:2012we}.  Additionally, of course, one needs to test the overall consistency of the 
total production cross-section with the SM prediction and here  currently  
CMS has  $\sigma/\sigma_{SM}=0.80\pm 0.22$ and ATLAS has $\sigma/\sigma_{SM}=1.2\pm 0.3$ ~\cite{July4}.
Important clues to new physics can emerge by looking at deviations of  the Higgs boson couplings from the standard model  predictions.  The possibility that CMS and ATLAS  data  may have  an excess in the diphoton channel has 
already led to  significant interest in the diphoton decay of the Higgs.  In the SM the Higgs boson can decay into photons via $W^+W^-$ and $t\bar t$ loops. The W-loop gives the larger contribution which is partially cancelled by the top-loop. 
To enhance the decay one could add extra charged particles in the loops  and various works  have appeared
along these lines (see, e.g.,  ~\cite{Carena:2012xa}) 
More data is needed to confirm the diphoton excess as the result could be due to QCD 
  uncertainties~\cite{Baglio:2012et}.\\

We discuss now the implications of the 125 GeV Higgs for supersymmetry (see also \cite{Zerwas}).
There are cogent arguments for the 125 GeV Higgs being the first evidence for supersymmetry at the LHC.
The argument assumes the following: (i) perturbative physics up to the grand unification scale; (ii)
 no large hierarchy problem. Item (i) disfavors composite Higgs models which require non-perturbative physics 
below the grand unification scale. Thus under  constraint (i) the standard model is a possible candidate 
since the SM is a renormalizable theory
and is valid up to the grand unification scale. However, the SM  gives large loop corrections to the Higgs mass,
so that at one loop one has
$ m^2_h = m_0^2 + O(\Lambda^2).$
Since  $\Lambda \sim O(10^{16})$ GeV  a large fine tuning is involved, i.e., one part in $10^{14}$. Thus the 
SM while satisfying (i) does not satisfy (ii).  On the other hand 
 SUGRA models  have all the merits of the SM, but are  free of the  
large hierarchy problem and  give perturbative physics up to the GUT scale. Here $\Lambda$ is essentially
replaced by the stop masses. \\

The second strong hint that the newly discovered particle  is a SUSY  Higgs is its mass.  
In the standard model the upper limit on the Higgs mass could be  much larger.
Using the condition that the weak interactions not become strong  one derives
roughly the upper bound~\cite{Lee:1977yc} $ M_H <  (8 \pi \sqrt 2 /3 G_F)^{1/2} \simeq 1 ~TeV$.
In SUSY, the Higgs quartic couplings are governed by gauge interactions and there is an upper limit which is much 
smaller, i.e.,  $M_H < 156$ ~GeV ~\cite{Kane:1992kq}
and in well motivated models such
as mSUGRA, GMSB, AMSB it is even smaller. Specifically in mSUGRA~\cite{can}
$M_H \leq 130 ~{\rm GeV}$~\cite{Akula:2011aa,Arbey:2012dq}.
It is important to realize that 
SUSY could  have been excluded or come under a shadow 
if the Higgs mass was found to be, say 200 GeV.  Thus it is very meaningful
that experimentally the Higgs mass ended up below the upper limit predicted in SUSY and more specifically
below the upper limit predicted in SUGRA models. In SUGRA 
 the dominant one loop contribution to the Higgs mass arises from the top/stop sector and is given by
$
\Delta m_h^2\simeq  {3m_t^4}/(2\pi^2 v^2) \ln (M_{\rm S}^2/m_t^2) 
+ {3 m_t^4}/(2 \pi^2 v^2)  ({X_t^2}/{M_{\rm S}^2} - {X_t^4}/(12 M_{\rm S}^4))~,
$
where  $v=246$~GeV, $M_{\rm S}$ is an average stop mass, and $X_t$ is given  by
  $X_t\equiv A_t - \mu \cot\beta~$ (for a review see~\cite{review}).
   The loop correction is maximized  when   
   $ X_t \sim \sqrt 6 M_{\rm S} ~.$ 
   Thus the 'large' Higgs mass, i.e., 125  GeV points to the SUSY scale being rather high. Indeed most of the allowed
   parameter space consistent with all the experimental constraints appears to lie on the 
   Hyperbolic Branch~\cite{Chan:1997bi,Feng:1999mn,Feldman:2011ud,Akula:2011jx}, and specifically on Focal
   Curves and on Focal Surfaces~\cite{Akula:2011jx}  (see Sec.4) which is another indication of the overall 
   SUSY scale being high.
   The 125 GeV Higgs boson also strongly constrains the sparticle 
   landscape~\cite{Feldman:2007zn}) (for a review see ~\cite{Nath:2010zj}).
   The analysis of the sparticle spectra under the constraints of the high Higgs boson mass 
   allows for light sparticle masses, i.e., the neutralino, the chargino, the gluino, the stau and the stop.
   The light gluino possibility has been extensively discussed in the literature (see. e.g., \cite{lowgluino,lowgluino2}).
 Thus a low lying gluino along with the neutralino, the chargino, and the stop appear prime candidates for
 discovery at the LHC, although a more quantitative analysis is needed correlated with the energy of the
 future runs at the LHC. The discovery potential for SUSY in mSUGRA
 at LHC at $\sqrt{s}=14$ TeV  and at high luminosity with 300 (3000) fb$^{-1}$ of integrated luminosity is discussed in ~\cite{Baer:2012vr}.

\section{GUTS and SUGRA GUTS}

Gauge symmetry based on $SO(10)$ provides a framework for unifying the $SU(3)_C\times SU(2)_L\times U(1)_Y$ gauge
groups and for unifying quarks and leptons in a single $16$--plet spinor representation.
Additionally, the 16--plet also contains a right--handed singlet state, which is needed to give mass to the neutrino via the seesaw mechanism. 
Supersymmetric $SO(10)$ models~\cite{georgi}
 have the added attraction that they predict correctly the unification of gauge couplings, and
solve the hierarchy problem by virtue of supersymmetry. 
 However, SUSY $SO(10)$ models, as usually constructed, have
two drawbacks, both related to the symmetry breaking sector.  
 First, two different mass scales  are involved in breaking of the GUT symmetry, one to
reduce the rank and the other to reduce the symmetry all the way to $SU(3)_C\times SU(2)_L\times U(1)_Y$. 
 Thus typically three types of Higgs fields are needed (for a review see ~\cite{Nath:2006ut})
    e.g., $16\oplus\overline {16}$ or $126\oplus\overline{126}$
 for rank reduction, and a $45$, $54$ or 
$210$ for breaking the symmetry down to the standard model symmetry, 
 and a 10-plet  for electroweak symmetry breaking.
Multiple  step breaking requires  additional assumptions  relating VEVs of  
different breakings  to explain gauge coupling unification of the electroweak and the strong interactions.
A single step breaking does not require such an assumption and  can be achieved with
 $144\oplus\overline{144}$~\cite{Babu:2005gx,Nath:2005bx} of Higgs fields.
 The reason for this is easily understood by looking at the decomposition of $144$   and $\overline{144}$ 
 multiplets 
   under  $SU(5)\times U(1)$. 
 Thus one has $\overline{144}  = \bar 5(3) + 5(7) + 10 (-1) + 15 (7) + 24(-5) + 40(-1) + \overline{45}(3)$.
Now the 24-plet carries a $U(1)$ quantum number and thus a VEV formation of it will
reduce the rank of the group as well as break $SU(5)$.
Additionally, one can obtain a pair of light Higgs doublets needed for electroweak symmetry breaking 
 from the  same irreducible $144\oplus\overline{144}$ Higgs multiplet.
Thus    $SO(10)$  reduces   to $SU(3)_C\times U(1)_{em}$  with just one  pair of  $<144\oplus\overline{144}>$
multiplets.\\

Second, GUT theories typically have the doublet-triplet problem, i.e., 
one must do an extreme fine--tuning at the level of one part in $10^{14}$ to get the Higgs doublets
of MSSM light, while  color-triplets  remain superheavy. Some possible solutions to the doublet-triplet problem include
 (i) missing  VEV mechanism where  $SO(10)$ breaks in the B-L direction, (ii) flipped $SU(5)\times U(1)$, (iii)  missing partner mechanism,
 (iv) orbifold GUTs. The  flipped $SU(5)\times U(1)$, 
 missing partner mechanism  and  the orbifold GUTs are rather  compelling in that some 
doublets  are forced to be massless. We will  focus on the missing partner mechanism~\cite{SU(5)Missingpartner}
 and discuss how it works in $SU(5)$ and then
discuss how one can extend to $SO(10)$. In SU(5) to obtain Higgs doublets which are naturally light one uses an array of light and heavy 
Higgs multiplets so that the heavy multplets consist of $50, \overline{50},75$ multiplets and the light multiplets consist of $5,\bar 5$, 
i.e., one has  mass terms for $75, 50, \overline{50}$ and no mass terms for $5+\bar 5$. 
The 75-plet breaks the GUT symmetry to $SU(3)\times SU(2)\times U(1)$.  The key element is that $50+\overline{50}$ have no
 doublet pairs (D) and only triplet/anti-triplet pairs (T). Thus  $50+ \overline{50}$ have   $0 D + 1 T$ while 
  $5+ \bar 5$ have   $1D + 1T$.
The Higgs triplets/anti-triplets of $50+\overline{50}$  mix with the Higgs triplets/anti-triplets 
of $5+\bar 5$ to become heavy.  This is accomplished via the superpotential
$W({\rm higgs})= W_0(75) +  M 50\otimes\overline{50} + \lambda_1 50\otimes 75\otimes\overline{5} + \lambda_2 \overline{50}\otimes 75\otimes 5 $.
 The doublets in $5\oplus\bar 5$ have nothing to pair up with and remain light.
 Next we extend the missing partner mechanism to $SO(10)$. Here one case has previously been worked out, i.e., 
 the case with the heavy sector consisting of 
 $126\oplus\overline{126}$ and the light sector  consisting of $10\oplus120$~\cite{SO(10)Missingpartner}. 
  In recent works 
 several other cases have been found~\cite{Babu:2011tw}. These models are anchored in  $126+\overline{126}$ or  $560\oplus \overline{560}$
 which play the role of  $50\oplus\overline{50}$. 
 Analysis of these larger representations require special tools which utilize oscillator techniques~\cite{ms,ns}.\\

 However, to connect the high scale physics to low energies we need to break supersymmetry. As is well known in global supersymmetry, it is 
 difficult to achieve the breaking of supersymmetry in a phenomenologically acceptable fashion. 
 To do so  one must consider local supersymmetry which involves gravity~ \cite{anz,fnf}.
 Thus one needs to construct supergravity grand unified models~\cite{can}, 
 coupled with radiative breaking of the electroweak symmetry (for a review see ~\cite{IbanezRoss}) 
 using renormalization group evolution of gauge, Yukawa and of soft terms ~\cite{Martin:1993zk}.  
 A broad class of models fall under this rubric. These include mSUGRA (CMSSM), 
 and SUGRA models with non-universalities in the Higgs sector and in the gaugino sector. 
  mSUGRA has the parameter space $ m_0, m_{1/2}, A_0, \tan\beta, {\rm sign}(\mu)$ where $m_0 $ is the universal 
  scalar mass, $m_{1/2}$ is the
  universal gaugino mass, $A_0$ is the universal trilinear coupling, $\tan\beta$ is the ratio $<H_2>/<H_1>$ where, $<H_2>$ gives  mass to the 
  up quarks and $<H_1>$ gives mass  to the down quarks and leptons and $\mu$ is the Higgs mixing parameter.  
   The  mSUGRA model assumes a flavor independent Kahler potential. However, the 
  nature of physics at the Planck scale is not fully understood. Thus it is useful to consider more 
  general Kahler potentials, 
  and indeed in string models and in D brane models  
  one often encounters more general Kahler potentials  (see, e.g., ~\cite{Kors:2003wf}). 
  Such choices give rise to non-universal supergravity models which we may label as
  NUSUGRA models. These include models with non-universal gaugino masses ~\cite{Chattopadhyay:2001mj}   
(see also ~\cite{DPRoy})  
  where the gaugino sector is characterized by the three
  parameters  $ \tilde m_1, \tilde m_2, \tilde m_3$ which characterize the gaugino masses for the gauge groups $U(1)$ , $SU(2)_L$ and $SU(3)_C$. 
  Similarly, one could have non-uinversality in the Higgs boson sector ~\cite{Nath:1997qm}  with different masses $ m_{H_1}, m_{H_2}$ for the Higgses $H_1$ and $H_2$, as well as non-universalities in the scalar sector.

\section{ Natural  TeV Size Scalars}
Radiative breaking of the electroweak symmetry naturally allows TeV size scalars without the necessity of excessive
fine tuning. To illustrate this phenomenon we consider the generic form of the radiative  breaking 
of the electroweak symmetry (REWSB)  which can be written in the form
$
 \mu^2  + \frac{1}{2}M_Z^2  = m^2_0  C_1+ A^2_0 C_2 +
 m^2_{\frac{1}{2}} C_3+ m_{\frac{1}{2}}
A_0 C_4+ \Delta \mu^2_{loop}.
$
With a redefinition of parameters one can write this equation in the form $\mu^2  + \frac{1}{2}M_Z^2  = m^2_0  C_1+ \bar A^2_0  C_2 +
  m^2_{{1}/{2}} \bar C_3+
  %m_{\frac{1}{2}} A_0 C_4+
   \Delta \mu^2_{loop}$ where 
   $\bar A_0 \equiv A_0 + ({C_4}/{2C_2}) m_{1/2}$  
   and $\bar C_3= C_3-{C_4^2}/{(4C_2)}$. 
To simplify matters one can go the renormatlization  group (RG) scale $Q$ where the loop contribution is minimized 
 (and  ideally vanishes). In this case we can use the tree part of the REWSB reliably. Typically $C_1, C_2, \bar C_3$ are 
all positive which implies that the soft parameters sit on the surface of an ellipsoid given by the REWSB equation.
Thus for fixed $\mu$ none of the soft parameters can get too large. One may call this the Ellipsoidal Branch~\cite{can}.
Suppose now that for certain choices of the inputs such as the top Yukawa, $\tan\beta$,  the gauge couplings
etc., that the co-efficient 
$C_1$ turns negative. In this case large value  of $m_0$ can occur due to the possibility of cancellation 
of the $C_1$ term 
with the rest of the
terms in REWSB. This is the Hyperbolic Branch~\cite{Chan:1997bi,Feng:1999mn,Feldman:2011ud,Akula:2011jx} 
which contains the Focal Point (HB/FP), Focal Curves (HB/FC), and Focal Surfaces (HB/FS).
The Focal Point corresponds to the case when $C_1$ vanishes~\cite{Feng:1999mn}, Focal Curves are 
when two parameters get large,
i.e., $m_0, A_0$ (HB/FC1) or  $m_0, m_{1/2}$ (HB/FC2), and Focal Surfaces (HB/FS) occur when all the three 
soft parameters $m_0, m_{1/2}, A_0$ get large while $\mu$ remains fixed and small. In all these cases one finds that large scalar masses exist. Using the LHC constraints the HB/FP region is seen to be depleted  \cite{Akula:2011jx} while most of the remaining 
parameter space appears to lie on Focal Curves and Focal Surfaces~\cite{Akula:2011jx,Akula:2011dd}. 
   
   \section{Dark Matter}
   In supergravity models the neutralino is the lightest R parity odd particle over most of the parameter space of the model 
   and with R parity it is a possible candidate for dark matter. The annihilation of dark matter proceeds via the $Z$ and
   the  Higgs 
   poles ($h^0, H^0, A^0$) in the s-channel and via the sfermion exchange in the t-channel. Thus the relic density is sensitive
 to the Higgs masses as well as to the masses of the sfermions. Similarly, the spin independent neutralino-proton cross-section 
 arises from the scattering of the neutralino from quarks in a nucleus and  the scattering  
 proceeds via  squark poles in the s-channel and via
 the $Z, h^0, H^0, A^0$ poles in the t-channel. Consequently 
 again the spin independent neutralino-proton cross-section 
 is sensitive to the scale of squark masses and to the Higgs masses. Thus one might ask as to the impact of the lightest Higgs
 mass at 125 GeV on dark matter. As discussed above, the direct implication of a 125 GeV Higgs mass is to raise the scale of supersymmetry.
 The larger scale leads to heavier sfermion masses which  implies that the relic density will be governed mostly by the
 s-channel Z and Higgs poles and by coannihilation. Analogously the spin independent neutralino-proton cross-section  will be largely governed
 by the t-channel Z and Higgs pole exchanges. These constraints lead to further limits on the allowed parameter 
 space of the model consistent with WMAP~\cite{wmap}. In this context we  mention that often the upper limit WMAP constraint is
 imposed so as to allow for the possibility of multicomponent dark matter~\cite{Feldman:2010wy}. An analysis of the Higgs constraint on 
the spin independent neutralino-proton cross-section was given in~\cite{Akula:2011aa,Akula:2012kk}
(see also ~\cite{Akula:2011dd})
 and it is found that most of the allowed parameter
space of mSUGRA model  lies between the current sensitivity of the XENON100 experiment~\cite{xenon} 
and the expected
sensitivity in the XENON-1T~\cite{futureXENON} and  SuperCDMS~\cite{futureSCDMS}. Thus the prospects for the discovery of dark matter look bright if the 
proposed sensitivities in the new generation of dark matter experiments can be reached.

\section{$g_{\mu}-2$ constraint}
The anomalous magnetic moment of the muon
$g_{\mu}-2$ is a sensitive test of new physics. In supersymetric models $g_{\mu}-2$ can receive important contributions from
supersymmetric loops. In the standard model electroweak contributions arise from the exchange of the W and Z  boson.
Analogously in supersymmtric models one has contributions arising from the exchange  of charginos and sneutrinos, and from the 
exchange of neutralinos and smuons~\cite{Yuan:1984ww}. If the sparticle spectrum is light, one finds that these contributions can be comparable
to or even exceed the standard model contribution. The experimental determination of the correction to $g_{\mu}-2$  or to $a_{\mu}$ defined by
$a_{\mu}=(g_{\mu}-2)/2$, i.e., $\delta a_{\mu} = a_{\mu}^{exp}-a_{\mu}^{SM}$ is sensitively dependent on the hadronic corrections in
the standard model. Thus there are two main estimations of the hadronic error: one using the $e^+e^-$ annihilation and the other
using $\tau$ decay. The analysis using $e^+e^-$ estimation of hadronic corrections gives~\cite{hoecker} 
 $\delta a_{\mu}= (28.7\pm 8.0)\times 10^{-10}$ which is a  $3.6\sigma$ deviation from the standard model, while the analysis using
 $\tau$ decay gives~\cite{hoecker} $\delta a_{\mu}= (19.5\pm 8.3)\times 10^{-10} ~(2.4\sigma)$ which is $2.4\sigma$ deviation from the standard model
 result. As discussed above  the high Higgs mass leads to a high scale for the  sfermions and with universal boundary conditions 
 on soft parameters they lead to a rather small correction to $a_{\mu}$. Thus there is  a tension between the $g_{\mu}-2$ experimental
 result and the high Higgs mass. Assuming the $g_{\mu}-2$ result does not undergo further oscillations, 
 this tension could be relieved if one assumes that part of the Higgs mass arises from sources other than from MSSM. 
 Thus extra matter could give additional  contribution to the Higgs mass~\cite{Babu:2008ge}
   effectively lowering the  component of the Higgs mass arising from  MSSM
 which would  allow an $a_{\mu}$ correction consistent with experiment. Similarly, the presence of an extra gauge group under which
 the Higgs  is charged would lead to a contribution from the D term to the Higgs mass again effectively lowering the MSSM Higgs
 mass.  Yet another possibility is that $g_{\mu}-2$ contribution arises from new physics other than SUSY. Thus, for example, a $Z'$ 
 which couples to the muon will produce a contribution to $a_{\mu}$. For, instance a $Z'$ arising from a  gauged $L_{\mu}-L_{\tau}$ 
 can have a significantly lower mass than the current experimental limits on a generic $Z'$ and can generate a correction to 
 $g_{\mu}-2$ of the right size ~\cite{Feldman:2010wy} (see the discussion in Sec.7).

\section{Cosmic coincidence}
An  interesting cosmic coincidences is the fact that the ratio of 
dark matter to baryonic matter is roughly 5, or more precisely
${\Omega_{DM}}/{ \Omega_B} = 4.99 \pm 0.20$~\cite{wmap}.
The fact that they are about the same size points to a possible common origin 
for them. 
The so called asymmetric dark matter (AsyDM) idea proposes that dark matter originated from 
baryonic matter in the early universe by the transfer of a net $B-L$ charge from the visible sector
to the dark sector  ~\cite{adm1} (For a review see ~\cite{Davoudiasl:2012uw}).
The implementation of this idea has two basic ingredients: first is to find a  mechanism for the transfer of $B-L$ 
from the visible sector  to the dark sector, and second to find a mechanism to deplete the symmetric component of dark 
matter produced by thermal processes. 
The standard procedure for the  transfer  of $B-L$  from the visible sector to the dark sector is  to consider
 an interactions of the type~\cite{admBL}
$M_a^{-n} O_{DM} O^{SM}_{asy}$ which is operative at temperatures $T_{int}$ so that 
$  T_{int} > (M_a^{2n} M_{Pl}^{-1})^{\frac{1}{2n-1}}, ~M_{Pl}= 2.4\times 10^{18} {\rm ~GeV}$
where $O^{SM}_{asy}$ is constituted of only standard model particles 
and $O_{DM}$ is constituted of dark matter fields and they  have opposite $B-L$ charges so that overall
the interaction is $B-L$ charge neutral. 
Concerning the depletion of the symmetric component of dark matter,  one needs a mechanism for an efficient 
annihilation of the thermally produced dark matter. \\

  The  early universe can be viewed as  a weakly interacting plasma in which each particle carries a chemical
 potential $\mu_i$. 
 In such a plasma the particle-anti-particle asymmetries are given by 
$ n_i- \bar n_i \simeq (g_i\beta T^3/6) (\mu_i  {\rm (fermi)}, 2 \mu_i{\rm  (bose)})$
where $g_i$ is the degrees of freedom, and $\beta=1/T$.  The chemical potentials are constrained by  
 (i)sphaleron interactions, (ii)  conservation of charge and hypercharge, (iii)  Yukawa and gauge interactions. 
When the transfer interaction is in equilibrium  (see, e.g., ~\cite{ht}), one can solve for the ratio $\Omega_{DM}/\Omega_B$ so that
$ {\Omega_{DM}}/{\Omega_B} =(X/B) (m_{DM}/m_B) \simeq 5$, 
where $X$ is the dark matter number density and $B$ is the baryon number density.
One can generate  a variety of models such as models where the visible sector is taken to be the standard model, 
 the two Higgs doublet model, or the MSSM~\cite{fnp}. 
For the MSSM many variations are possible  depending on the scale of sparticle masses. 
For each model  there are various interactions
that allow a transfer of the $B-L$ asymmetry from
the standard model sector to the dark matter sector.
The relic density in AsyDM will have two components:  a thermal and a non-thermal component. Thus one may 
write the total as a sum
$\Omega_{\rm DM} = \Omega_{\rm DM}^{\rm asy} + \Omega_{\rm DM}^{\rm sym}$.
For AsyDM to work we need  $ \Omega_{\rm DM}^{\rm sym} <<  \Omega_{\rm DM}^{\rm asy}$.
Thus we need an efficient mechanism for the annihilation of thermally produced  dark matter.
We  accomplish this via the exchange of a gauge field using
the \st formalism 
where the gauge field couples to $L_{\mu}- L_{\tau}$~\cite{fnp}.
In the unitary gauge the massive vector boson field will be called $Z'$ and its
  interaction  with fermions  is given by
$
L_{\rm int} = Q^{\psi} g_C \bar{\psi}  \gamma^{\mu} \psi Z'_{\mu}  +  Q^f g_C \bar f \gamma^{\mu} f {Z'_{\mu}}~, ~f=\mu, \tau.
$
where $f$ runs over $\mu$ and $\tau$ families and  $Q_C^{\mu}=-Q^{\tau}_C$.
The LEP constraints on the $M_{Z'}$ mass are not valid since $Z'$ does not couple with the first generation
leptons. This result also holds at the loop level to a good approximation. 
 The strongest constraint comes from $g_{\mu}-2$.
Thus the correction to $g_{\mu}-2$ arising from the exchange of $Z'$ boson associated with the $L_{\mu}-L_{\tau}$ symmetry is given 
by
$
\Delta (g_{\mu}-2) = (\frac{1}{2} g_C Q_C^{\mu})^2 {m_{\mu}^2}/{(6\pi^2 M_{Z'}^2)}.
$
Imposing the constraints $\Delta a_{\mu} = \Delta (g_{\mu} -2)/2\leq 3 \times 10^{-9}$ one finds the 
restriction 
$
M_{Z'}/(g_CQ_C^{\mu}) \geq 90~{\rm GeV}.
$
The above constraint allows for a low lying $Z'$ which couples only to muons and taus and allows for a rapid
annihilation of symmetric dark  matter via the $Z'$ pole.  \\

 To obtain relic densities at current temperatures for $\psi$ and $\bar \psi$ one must solve the Boltzman
 equations in the presence of asymmetries. 
The Boltzmann equations obeyed by
$f_{\psi}$ and  $f_{\bar\psi}$  take the form
$
  {d f_{\psi}}/{d x} = \alpha \langle \sigma v\rangle (f_{\psi} f_{\bar \psi} - f^{\rm eq}_{\psi} f^{\rm eq}_{\bar \psi})\,,
$
$
  {d f_{\bar\psi}}/{d x}  = \alpha \langle \sigma v \rangle (f_{\psi} f_{\bar \psi} - f^{\rm eq}_{\psi} f^{\rm eq}_{\bar \psi})\,,
$
where $x=k_BT/m_{\psi}$ and $f_{\psi} = { n_{\psi}}/{ h T^3}, ~~f_{\bar\psi} = { n_{\bar\psi}}/{ h T^3}$
where h is the entropy degrees of freedom.
One finds that 
$ \gamma =  f_{\psi}  -f_{\bar\psi},$   is a constant independent of temperature. 
 The relic densities for $\psi$ and $\bar \psi$ are then given by
 $
{\Omega_{\psi} h_0^2}/{(\Omega_{\psi}h_0^2)_{\xi=0}} \simeq
\xi {J(x_f)}/{(1 -  exp(- \xi J(x_f))}\ \to 1$ as $\xi\to 0$,
where $\xi = \gamma C$ and where $C$ is a numerical constant and $J(x_f)  \equiv \int_{x_0}^{x_f} \langle \sigma v \rangle \, d x$.
Similarly,
$
{\Omega_{\bar \psi} h_0^2}/{(\Omega_{\psi} h_0^2)}\simeq exp(- \xi J(x_f))\  \to 1$ as  $\xi\to 0$.
We need to show that 
$\Omega_{\bar \psi} h_0^2 << (\Omega h_0^2)_{\rm  WMAP}$ and that  $\Omega_{\psi} h_0^2$
is the major component of WMAP.
This has been exhibited explicitly in ~\cite{fnp}. 
We discuss now collider implications of the model.
In a muon collider there would be final states with muons and taus and their neutrinos but no 
$e^+e^-$ final states providing a smoking gun signature for the model ~\cite{fnp}.   The analysis is done
including one loop corrections arising from the first and the second generation leptons in the loop. 
One can carry out a direct extension of AsyDM to the supersymmetric case. 
The basic interaction responsible for the asymmetry has the form 
$
W_{\rm asy} = M_a^{-n} O_{DM} O^{\rm mssm}_{\rm asy}~.
$
In general there are many possibilities for the operators $O^{\rm mssm}_{\rm asy}$ such as   
$LH_2, LLE^C, QLD^C, U^CD^CD^C$
 or any product thereof. Obviously $O_{\rm DM}$ 
will carry the opposite quantum numbers to those of $O^{\rm mssm}_{\rm asy}$.  
In this case there are two dark matter particles, i.e., $\psi$ and the neutralino $\tilde \chi^0$ and here the total relic density is
$
\Omega_{\rm DM} = \Omega_{\rm DM}^{\rm asy} + \Omega_{\rm DM}^{\rm sym}  + \Omega_{\tilde\chi^0}\,,
$
where $\Omega_{\tilde \chi^0}$ is the relic density from the neutralino. One must show that the neutralino
contribution is subdominant, i.e., it is no more than 10\% of the WMAP value.
An interesting question is if a subdominant neutralino is detectable. This appears to be the case
as demonstrated in ~\cite{fnp}.

\section{R parity and Proton Stability}
  R parity is needed to get rid of baryon and lepton number  violating dimension 4 operators which can generate unacceptably fast 
  proton decay.  Within MSSM R parity is ad hoc. Further,  R parity as a global symmetry is not desirable since it can be broken by  wormhole effects.   This problem can be evaded if a model possesses  a  gauge symmetry so that  $R$ parity arises as a discrete remnant of this gauge symmetry (see, e.g., ~\cite{Martin:1992mq}). Since $R=(-1)^{2S + 3 (B-L)}$ the obvious extended symmetry is  $ SU(3)_C\times SU(2)_L\times U(1)_Y\times U(1)_{B-L}$.
In this case the $U(1)_{B-L}$ gauge symmetry will forbid R parity violating interactions such as  
$QLd^c, LLe^c, u^cd^cd^c, L\bar H$.
 Now an unbroken  $U(1)_{B-L}$ gauge symmetry is undesirable since there would be associated with it a massless gauge boson which
 can generate an unacceptable long range force and thus one must generate a mass for the $B-L$ gauge boson. 
 If the $B-L$ gauge symmetry is broken spontaneously, R parity is no longer guaranteed. Specifically R parity is 
protected if $3(B-L)$ is an even integer  but is  not protected if $3(B-L)$ is an odd integer.  
 To illustrate this point in a specific example one may consider   
 an extension of MSSM with a $U(1)_{B-L}$ symmetry which includes three right handed neutrinos 
    fields $\nu^c$ for anomaly cancellation. The extended superpotential in this case is
 ${\cal W}= {\cal W}_{\rm MSSM} \ + \ h_\nu \ L H_u \nu^c  + h_{\nu^c} \nu^c\nu^c \Phi + \mu_{\Phi} \Phi \bar \Phi $ where 
 the new fields are assigned the $B-L$ quantum numbers as $ (\nu^c, \Phi,  \bar \Phi): (-1, -2, 2) $.  
Since the $B-L$ of the $\Phi$ field is even, a VEV growth for it will not violate R parity. However, the VEV growth for 
$\nu^c$ for which $B-L$ is odd, will violate  R parity.  It turns out that the latter possibility can be realized 
in radiative breaking. This is so because 
    the beta functions
  due to the coupling of the $\Phi$ and $\nu^c$ can turn the mass of  $\nu^c$ tachyonic which leads
  to a VEV growth for $\nu^c$ and a violation of R parity~\cite{Barger:2008wn,Perez}.\\
 
 A violation of  R parity by the $\nu^c$ VEV growth can be avoided if one uses the \st mechanism ~\cite{kn12,FLNPRL} (see also ~\cite{Ringwald})  for the mass growth of the $B-L$ gauge boson rather than spontaneous breaking. Specifically
 let us  assume a \st mechanism for the $B-L$ gauge boson mass growth with the additional assumptions 
$\left <\tilde q\right >=0, ~\left<\tilde e_L\right>=0=  \left<\tilde e^c\right> $ which are just the conditions for charge 
conservation. Since $\tilde \nu_L$ and $\tilde e_L$ belong to the same $SU(2)_L$ multiplet, the 
vanishing of $\left<\tilde e_L\right>$ also implies the vanishing of $<\tilde \nu_L>$, i.e., 
$<\tilde \nu_L>=0$ since  the RG evolution of $M_{\tilde e_L}$ and of $M_{\tilde \nu_L}$ 
are very similar. It then follows after integrating over the remaining \st fields that the scalar potential takes the 
form 
$
V_{\nu^c}= M_{\tilde\nu^c}^2  \tilde \nu^{c\dag}\tilde \nu^c   + 
(g^2_{BL} M_{\rho}^2)/[2 (M_{BL}^2 + M_{\rho}^2)] (\tilde \nu^{c\dag} \tilde\nu^c)^2.$
In this case there are no beta functions to turn $M_{\tilde\nu^c}^2$ negative in the renormalization group evolution. 
As a consequently the potential cannot support spontaneous breaking to generate a VEV of 
$\tilde\nu^c$ and  $\left<\tilde \nu^c\right>=0$.
 Thus with the \st mechanism the $B-L$ gauge boson gains a mass but R parity remains
 unbroken. \\

 We discuss now briefly proton decay.  While R parity can eliminate  baryon and lepton number  violating dimension 4 operators,
 baryon and lepton number  violating  dimension five and dimension six  operators do exist in unified models. The
 baryon and lepton number  violating dimension five operators arise from color higgsino exchange  while the baryon and lepton number  violating 
 dimension six operators arise from the exchange of lepto-quarks.  The baryon and lepton number  violating dimension six
 operators lead to the dominant proton decay mode  $p\to e^+\pi^0$ and estimates give a lifetime
 which could be in the $10^{35}$ yrs range (For a recent review see  ~\cite{Hewett:2012ns}). The current 
 experimental limit from SuperKamiokande  for this mode is   $\tau(p\to e^+\pi^0) > 1.4\times 10^{34} {\rm yrs}$.
 It is expected that at Hyper-Kamiokande ~\cite{Abe:2011ts} one will reach the limit $\tau(p\to e^+\pi^0) > 1\times 10^{35} {\rm yrs}$,
 and thus there is a chance for the observation of the $e^+\pi^0$ mode.  Proton decay from
  baryon and lepton number  violating dimension five operators is more model dependent. Thus the baryon and lepton number  violating dimension five 
  operators   must be dressed  by chargino, gluino, and neutralino exchanges to produce baryon and lepton number  violating dimension six
  operators responsible for proton decay.  Since the sparticle spectrum enters in the dressing loops, the 
  proton lifetime is 
   dependent on the nature of the sparticle spectrum as well on CP phases  (for a review
  see ~\cite{Ibrahim:2007fb}) and thus proton decay 
  provides a test of models of  strings and branes via proton decay 
  branching ratios (see, e.g., ~\cite{Arnowitt:1993pd},\cite{Nath:2006ut}).    
   Specifically a light sparticle spectrum will  lead to a shorter proton decay while a heavier  spectrum will lead to
  a longer proton decay lifetime. The current experimental limit on p decay from SuperKamiokande
   gives $\tau(p\to \bar \mu K^+) > 4\times 10^{33} {\rm yrs}$ which puts severe constraints on supersymmetric 
   models. These constraints can be relieved either by GUT models which invoke a cancellation mechanism for
   the baryon and lepton number  violating dimension five operators or a heavy sparticle spectrum. The data from the large hadron 
   collider which indicates that the scale of SUSY breaking for the scalar sector may be large helps stabilize
   the proton against too fast a decay from dimension five operators.   It is expected that 
   Hyper-Kamiokande ~\cite{Abe:2011ts}   will 
   reach a sensitivity of $  2\times 10^{34} {\rm yrs}$ for the $\tau(p\to \bar \mu K^+)$ mode. In any case
   proton decay from dimension five operators could become visible even with modest increase in 
   sensitivity for the detection of this mode in the future.

\section{Conclusion}
We have given a brief summary of the current status of supergravity grand unification. A very significant 
constraint on the model arises  from the recent discovery of the Higgs boson by the ATLAS and CMS 
detectors and the determination that the mass of the Higgs is high, i.e., around 125 GeV.  It has been
known for some time that in mSUGRA the mass of  the Higgs must lie below 130 GeV, and thus
it is interesting that the mass of the Higgs as determined by experiment lies below the predicted limit.
However, the high mass of the Higgs  implies typically that the average mass of the stops must be
significantly large and further that the trilinear coupling $A_0$ typically must be large. These
results imply that the part of the parameter space of mSUGRA which gives rise to the large Higgs mass
would lie on the Hyperbolic Branch of radiative breaking of the electroweak symmetry and more
specifically on Focal Curves or Focal Surfaces. A theoretical analysis using the Higgs mass as  a constraint 
also implies that there is a significant region of the parameter space  consistent with all experimental
constraints where the gaugino masses are relatively light. Additionally in some restricted regions of the
parameter space on may also have relatively light stops and staus. Thus several light sparticles appear to be 
prime candidates for discovery at the LHC. These include the neutralino, the chargino, the gluino, and 
the stop. Another implication of  the heavy Higgs mass pertains to neutralino dark matter. 
The Higgs mass constraint
 significantly  restricts the allowed range of the    spin independent neutralino-proton 
cross section. Interestingly most of the allowed range lies between the current sensitivity of the
XENON100 and the expected sensitivity of XENON-1T and SuperCDMS experiments. 
This result is very encouraging for the discovery of supersymmetric dark matter. 
The issue of the $g_{\mu}-2$  constraint was also discussed. The excess of $g_{\mu}-2$ over the standard 
model result is dependent on the hadronic corrections. The $e^+e^-$ annihilation data and the $\tau$ decay
data give somewhat different results for this quantity leading to either $3.6 \sigma$ or $2.4\sigma$ 
discrepancy.  If the deviation from the standard model stays there would be tension between the
$g_{\mu}-2$ result and the high Higgs mass in models with universal boundary conditions  on the 
soft parameters. This tension can be relieved in several ways, such as by having part of the Higgs
mass arise from extra matter at the loop level, or from D terms if there is an extra gauge group 
under which the Higgs is charged. Alternately flavored supergravity models with non-universalities in the
flavor sector could account for the discrepancy. \\

Also discussed was the status of grand unification. Currently $SO(10)$ is the favored  group
for the unification of the electroweak and the strong interactions. $SO(10)$ has the  advantage 
over $SU(5)$ in that it unifies a full generation of quarks and leptons in a single 16-plet representation. However, 
the current models based on $SO(10)$ suffer from a couple of drawbacks. The first one concerns
the breaking of the GUT group. Here in conventional models  more than one scale enters in the 
breaking of  the group. Thus one scale enters in reducing  the rank and the other to break the symmetry
all the way down to the standard model gauge group. This drawback can be overcome if one
considers $144+\overline{144}$ of Higgs whose VEV formation can break the group down to the standard
model gauge group in one step.  A second draw back which is a generic problem in GUT models concerns
the doublet-triple splitting problem.
A new class of  $SO(10)$ 
models using $560+\overline{560}$ of Higgs fields not only break the gauge symmetry at one scale
they  also solve the doublet-triplet problem by the missing partner mechanism.  Further, work 
using these models is thus desirable.  The baryon and lepton number  violating dimension five and dimension six operatos which 
give rise to proton decay and the  prospects for the discovery of proton decay  in future 
experiments were discussed. 
Finally we discussed the so called  cosmic coincidence which pertains to the fact that the 
dark matter and the baryonic matter are in the ratio $\sim 5:1$. The realization of this possibility in the context of
superymmetry was discussed in a multicomponent dark matter framework consisting of a Dirac fermion
(carrying $B-L$ charge) and a Majorana fermion (the neutralino). \\

\noindent
{\bf Acknowledgments}\\
This research is supported in part by the U.S. National Science Foundation (NSF) grants
PHY-0757959 and PHY-0969739  and through XSEDE under grant number TG-PHY110015,
and NERSC grant  DE-AC02-05CH11231.\\


\begin{thebibliography}{999}

\bibitem{ATLAShiggs}
 F.~Gianotti [on behalf of ATLAS],
 ``Update on the Standard Model Higgs searches in ATLAS", joint CMS/ATLAS seminar, December 13, 2011; [ATLAS Collaboration] ATLAS-CONF-2011-163.
  [ATLAS Collaboration],
  %``Combined search for the Standard Model Higgs boson using up to 4.9 fb-1 of pp collision data at sqrt(s) = 7 TeV with the ATLAS detector at the LHC,''
  arXiv:1202.1408 [hep-ex].
  %%CITATION = ARXIV:1202.1408;%%

\bibitem{CMShiggs}
 G.~Tonelli [on behalf of CMS],
 ``Update on the Standard Model Higgs searches in CMS", joint CMS/ATLAS seminar, December 13, 2011; [CMS Collaboration] CMS PAS HIG-11-032.

\bibitem{HiggsBoson}
 F.~Englert and R.~Brout,
  %``Broken Symmetry and the Mass of Gauge Vector Mesons,''
Phys.\ Rev.\ Lett.\ \ {\bf 13}, 321  (1964);
%%CITATION = PRLTA,13,321;%%
  P.~W.~Higgs,
  %``Broken symmetries, massless particles and gauge fields,''
Phys.\ Lett.\ \ {\bf 12}, 132  (1964).
%%CITATION = PHLTA,12,132;%%
 % %``Broken Symmetries and the Masses of Gauge Bosons,''
Phys.\ Rev.\ Lett.\ \ {\bf 13}, 508  (1964);
%%CITATION = PRLTA,13,508;%%
  G.~S.~Guralnik, C.~R.~Hagen and T.~W.~B.~Kibble,
  %``Global Conservation Laws and Massless Particles,''
Phys.\ Rev.\ Lett.\ \ {\bf 13}, 585  (1964).
%%CITATION = PRLTA,13,585;%% 

\bibitem{Akula:2011aa} 
  S.~Akula, B.~Altunkaynak, D.~Feldman, P.~Nath and G.~Peim,
  %``Higgs Boson Mass Predictions in SUGRA Unification, Recent LHC-7 Results, and Dark Matter,''
  Phys.\ Rev.\ D {\bf 85}, 075001 (2012).
%  [arXiv:1112.3645 [hep-ph]].
  %%CITATION = ARXIV:1112.3645;%%

\bibitem{higgs_7tev1}
  H.~Baer \etal%, V.~Barger and A.~Mustafayev,
  %``Implications of a 125 GeV Higgs scalar for LHC SUSY and neutralino dark matter searches,''
  Phys.\ Rev.\ D {\bf 85}, 075010 (2012);
  %[arXiv:1112.3017 [hep-ph]].
  %%CITATION = ARXIV:1112.3017;%%
    A.~Arbey \etal%, M.~Battaglia, A.~Djouadi, F.~Mahmoudi and J.~Quevillon,
  %``Implications of a 125 GeV Higgs for supersymmetric models,''
  Phys.\ Lett.\ B {\bf 708}, 162 (2012);
%  [arXiv:1112.3028 [hep-ph]].
  %%CITATION = ARXIV:1112.3028;%%
    J.~L.~Feng \etal%, K.~T.~Matchev and D.~Sanford,
  %``Focus Point Supersymmetry Redux,''
  Phys.\ Rev.\ D {\bf 85}, 075007 (2012);
%  [arXiv:1112.3021 [hep-ph]].
  %%CITATION = ARXIV:1112.3021;%%
   M.~Carena \etal%, S.~Gori, N.~R.~Shah and C.~E.~M.~Wagner,
  %``A 125 GeV SM-like Higgs in the MSSM and the $\gamma \gamma$ rate,''
  JHEP {\bf 1203}, 014 (2012);
%  [arXiv:1112.3336 [hep-ph]].
  %%CITATION = ARXIV:1112.3336;%%
    S.~Heinemeyer \etal%, O.~Stal and G.~Weiglein,
  %``Interpreting the LHC Higgs Search Results in the MSSM,''
  Phys.\ Lett.\ B {\bf 710}, 201 (2012);
%  [arXiv:1112.3026 [hep-ph]].
  %%CITATION = ARXIV:1112.3026;%%
    P.~Draper \etal%, P.~Meade, M.~Reece and D.~Shih,
  %``Implications of a 125 GeV Higgs for the MSSM and Low-Scale SUSY Breaking,''
  Phys.\ Rev.\ D {\bf 85}, 095007 (2012);
%  [arXiv:1112.3068 [hep-ph]].
  %%CITATION = ARXIV:1112.3068;%%
      J.~Ellis   and K.~A.~Olive,
%  %``Revisiting the Higgs Mass and Dark Matter in the CMSSM,''
  Eur.\ Phys.\ J.\ C {\bf 72}, 2005 (2012); 
%  %%CITATION = ARXIV:1202.3262;%%  
    O.~Buchmueller \etal%, R.~Cavanaugh, A.~De Roeck, M.~J.~Dolan, J.~R.~Ellis, H.~Flacher, S.~Heinemeyer and G.~Isidori {\it et al.},
  %``Higgs and Supersymmetry,''
 arXiv:1112.3564 [hep-ph];
  %%CITATION = ARXIV:1112.3564;%%
    U.~Ellwanger,
  %``A Higgs boson near 125 GeV with enhanced di-photon signal in the NMSSM,''
  JHEP {\bf 1203}, 044 (2012).
%  [arXiv:1112.3548 [hep-ph]].

\bibitem{higgs_7tev2} 
    M.~Kadastik \etal%, K.~Kannike, A.~Racioppi and M.~Raidal,
  %``Implications of the 125 GeV Higgs boson for scalar dark matter and for the CMSSM phenomenology,''
  JHEP {\bf 1205}, 061 (2012);
%  [arXiv:1112.3647 [hep-ph]].
  %%CITATION = ARXIV:1112.3647;%%
    J.~Cao \etal%, Z.~Heng, D.~Li and J.~M.~Yang,
  %``Current experimental constraints on the lightest Higgs boson mass in the constrained MSSM,''
  Phys.\ Lett.\ B {\bf 710}, 665 (2012);
%  [arXiv:1112.4391 [hep-ph]].
  %%CITATION = ARXIV:1112.4391;%%
  B.~Batell \etal%, S.~Gori and L.~-T.~Wang,
  %``Exploring the Higgs Portal with 10/fb at the LHC,''
   arXiv:1112.5180 [hep-ph];
  %%CITATION = ARXIV:1112.5180;%%  
     L.~J.~Hall \etal%, D.~Pinner and J.~T.~Ruderman,
  %``A Natural SUSY Higgs Near 126 GeV,''
  JHEP {\bf 1204}, 131 (2012);
%  [arXiv:1112.2703 [hep-ph]].
  %%CITATION = ARXIV:1112.2703;%%
    J.~F.~Gunion \etal%, Y.~Jiang and S.~Kraml,
  %``The Constrained NMSSM and Higgs near 125 GeV,''
  Phys.\ Lett.\ B {\bf 710}, 454 (2012);
%  [arXiv:1201.0982 [hep-ph]].
  %%CITATION = ARXIV:1201.0982;%%
 P.~Fileviez Perez,
  %``SUSY Spectrum and the Higgs Mass in the BLMSSM,''
  Phys.\ Lett.\ B {\bf 711}, 353 (2012);
  %%CITATION = ARXIV:1201.1501;%%
    S.~F.~King \etal%, M.~Muhlleitner and R.~Nevzorov,
  %``NMSSM Higgs Benchmarks Near 125 GeV,''
  Nucl.\ Phys.\ B {\bf 860}, 207 (2012);
%  [arXiv:1201.2671 [hep-ph]].
  %%CITATION = ARXIV:1201.2671;%%
    J.~Cao \etal%, Z.~Heng, J.~M.~Yang, Y.~Zhang and J.~Zhu,
  %``A SM-like Higgs near 125 GeV in low energy SUSY: a comparative study for MSSM and NMSSM,''
  JHEP {\bf 1203}, 086 (2012);
%  [arXiv:1202.5821 [hep-ph]].
  %%CITATION = ARXIV:1202.5821;%%
  C.~-F.~Chang \etal%, K.~Cheung, Y.~-C.~Lin and T.~-C.~Yuan,
  %``Mimicking the Standard Model Higgs Boson in UMSSM,''
  JHEP {\bf 1206}, 128 (2012);
  %%CITATION = ARXIV:1202.0054;%%
   L.~Aparicio \etal%, D.~G.~Cerdeno and L.~E.~Ibanez,
  %``A 119-125 GeV Higgs from a string derived slice of the CMSSM,''
  JHEP {\bf 1204}, 126 (2012);
%  [arXiv:1202.0822 [hep-ph]].
  %%CITATION = ARXIV:1202.0822;%%
  H.~Baer \etal%, V.~Barger and A.~Mustafayev,
  %``Neutralino dark matter in mSUGRA/CMSSM with a 125 GeV light Higgs scalar,''
  JHEP {\bf 1205}, 091 (2012);
  %%CITATION = ARXIV:1202.4038;%%
  D.~Ghosh \etal%, M.~Guchait and D.~Sengupta,
  %``Higgs signal in Chargino-Neutralino production at the LHC,''
  arXiv:1202.4937 [hep-ph].
  %%CITATION = ARXIV:1202.4937;%%
N.~Desai \etal%, B.~Mukhopadhyaya and S.~Niyogi,
  %``Constraints on invisible Higgs decay in MSSM in the light of diphoton rates from the LHC,''
  arXiv:1202.5190 [hep-ph];
  %%CITATION = ARXIV:1202.5190;%%
      D.~Carmi \etal%, A.~Falkowski, E.~Kuflik and T.~Volansky,
  %``Interpreting LHC Higgs Results from Natural New Physics Perspective,''
  arXiv:1202.3144 [hep-ph];
  %%CITATION = ARXIV:1202.3144;%%
  L.~Maiani \etal%, A.~D.~Polosa and V.~Riquer,
  %``Probing Minimal Supersymmetry at the LHC with the Higgs Boson Masses,''
  arXiv:1202.5998 [hep-ph];
  %%CITATION = ARXIV:1202.5998;%%
  T.~Cheng \etal%, J.~Li, T.~Li, D.~V.~Nanopoulos and C.~Tong,
  %``Electroweak Supersymmetry around the Electroweak Scale,''
  arXiv:1202.6088 [hep-ph];
  %%CITATION = ARXIV:1202.6088;%%
  J.~Cao \etal%, Z.~Heng, J.~M.~Yang and J.~Zhu,
  %``Higgs decay to dark matter in low energy SUSY: is it detectable at the LHC ?,''
  JHEP {\bf 1206}, 145 (2012);
  %%CITATION = ARXIV:1203.0694;%%
  F.~Jegerlehner,
  %``Implications of low and high energy measurements on SUSY models,''
  arXiv:1203.0806 [hep-ph];
  %%CITATION = ARXIV:1203.0806;%%
  N.~D.~Christensen \etal%, T.~Han and S.~Su,
  %``MSSM Higgs Bosons at The LHC,''
  arXiv:1203.3207 [hep-ph];
  %%CITATION = ARXIV:1203.3207;%%
  A.~Choudhury and A.~Datta,
  %``Many faces of low mass neutralino dark matter in the unconstrained MSSM, LHC data and new signals,''
  JHEP {\bf 1206}, 006 (2012);
  %%CITATION = ARXIV:1203.4106;%%
  B.~S.~Acharya \etal%, G.~Kane and P.~Kumar,
  %``Compactified String Theories -- Generic Predictions for Particle Physics,''
  Int.\ J.\ Mod.\ Phys.\ A {\bf 27}, 1230012 (2012);
  %%CITATION = ARXIV:1204.2795;%% 
  M.~A.~Ajaib \etal%, I.~Gogoladze, F.~Nasir and Q.~Shafi,
  %``Revisiting mGMSB in light of a 125 GeV Higgs,''
  arXiv:1204.2856 [hep-ph];
  %%CITATION = ARXIV:1204.2856;%%
    J.~Ellis and T.~You,
  %``Global Analysis of Experimental Constraints on a Possible Higgs-Like Particle with Mass ~ 125 GeV,''
  arXiv:1204.0464 [hep-ph];
  %%CITATION = ARXIV:1204.0464;%%
   C.~Balazs \etal%, A.~Buckley, D.~Carter, B.~Farmer and M.~White,
  %``Should we still believe in constrained supersymmetry?,''
  arXiv:1205.1568 [hep-ph];
  %%CITATION = ARXIV:1205.1568;%%
   M.~Badziak, E.~Dudas, M.~Olechowski and S.~Pokorski,
  %``Inverted sfermion mass hierarchy and the Higgs boson mass in the MSSM,''
  JHEP {\bf 1207}, 155 (2012)
  [arXiv:1205.1675 [hep-ph]];
  %%CITATION = ARXIV:1205.1675;%%  
  J.~L.~Feng and D.~Sanford,
  %``A Natural 125 GeV Higgs Boson in the MSSM from Focus Point Supersymmetry with A-Terms,''
  arXiv:1205.2372 [hep-ph];
  %%CITATION = ARXIV:1205.2372;%%
  N.~Okada,
  %``SuperWIMP dark matter and 125 GeV Higgs boson in the minimal GMSB,''
  arXiv:1205.5826 [hep-ph];
  %%CITATION = ARXIV:1205.5826;%%
  M.~Carena \etal%, S.~Gori, N.~R.~Shah, C.~E.~M.~Wagner and L.~-T.~Wang,
  %``Light Stau Phenomenology and the Higgs \gamma\gamma Rate,''
  arXiv:1205.5842 [hep-ph];
  %%CITATION = ARXIV:1205.5842;%%
  E.~Dudas \etal%, Y.~Mambrini, A.~Mustafayev and K.~A.~Olive,
  %``Relating the CMSSM and SUGRA models with GUT scale and Super-GUT scale Supersymmetry Breaking,''
  arXiv:1205.5988 [hep-ph];
  %%CITATION = ARXIV:1205.5988;%%
  A.~Fowlie \etal%, M.~Kazana, K.~Kowalska, S.~Munir, L.~Roszkowski, E.~M.~Sessolo, S.~Trojanowski and Y.~-L.~S.~Tsai,
  %``The CMSSM Favoring New Territories: The Impact of New LHC Limits and a 125 GeV Higgs,''
  arXiv:1206.0264 [hep-ph];
  %%CITATION = ARXIV:1206.0264;%%
  M.~Hirsch \etal%, W.~Porod, L.~Reichert and F.~Staub,
  %``Phenomenology of the minimal supersymmetric $U(1)_{B-L}\times U(1)_R$ extension of the standard model,''
  arXiv:1206.3516 [hep-ph];
  %%CITATION = ARXIV:1206.3516;%%
  R.~M.~Chatterjee \etal%, M.~Guchait and D.~Sengupta,
  %``Probing Supersymmetry using Event Shape variables at 8 TeV LHC,''
  arXiv:1206.5770 [hep-ph]; 
  %%CITATION = ARXIV:1206.5770;%%
M.~W.~Cahill-Rowley \etal%, J.~L.~Hewett, A.~Ismail and T.~G.~Rizzo,
  %``The Higgs Sector and Fine-Tuning in the pMSSM,''
  arXiv:1206.5800 [hep-ph];
  %%CITATION = ARXIV:1206.5800;%%
    A.~Albaid and K.~S.~Babu,
  %``Higgs boson of mass 125 GeV in GMSB models with messenger-matter mixing,''
  arXiv:1207.1014 [hep-ph];
  %%CITATION = ARXIV:1207.1014;%%
 C.~Boehm, J.~Da Silva, A.~Mazumdar and E.~Pukartas,
  %``Probing the Supersymmetric Inflaton and Dark Matter link via the CMB, LHC and XENON1T experiments,''
  arXiv:1205.2815 [hep-ph].
  %%CITATION = ARXIV:1205.2815;%%



\bibitem{tevatron}
%  , {\it et al.}  
[The TeVatron],
  %``Updated Combination of CDF and D0 Searches for Standard Model Higgs Boson Production with up to 10.0 fb-1 of Data,''
  arXiv:1207.0449 [hep-ex].
  %%CITATION = ARXIV:1207.0449;%%

\bibitem{July4}
J.~Incandela [on behalf of CMS],
``Status of the CMS SM Higgs Search", and 
F.~Gianotti [on behalf of ATLAS],
``Status of the SM Higgs Search in ATLAS",  joint CMS/ATLAS seminars at CERN, July 4, 2012;
CMS Collaboration,  CMS-PAS-HIG-12-020;  ~ ATLAS Collaboration, ATLAS-CONF-2012-093.  

\bibitem{Peskin:2012we} 
  M.~E.~Peskin,
  %``Comparison of LHC and ILC Capabilities for Higgs Boson Coupling Measurements,''
  arXiv:1207.2516 [hep-ph].
  %%CITATION = ARXIV:1207.2516;%%

\bibitem{Carena:2012xa} 
  M.~Carena, I.~Low and C.~E.~M.~Wagner,
  %``Implications of a Modified Higgs to Diphoton Decay Width,''
  arXiv:1206.1082 [hep-ph].
  %%CITATION = ARXIV:1206.1082;%%

\bibitem{Baglio:2012et} 
  J.~Baglio, A.~Djouadi and R.~M.~Godbole,
  %``The apparent excess in the Higgs to di-photon rate at the LHC: New Physics or QCD uncertainties?,''
  arXiv:1207.1451 [hep-ph].
  %%CITATION = ARXIV:1207.1451;%%

\bibitem{Zerwas}
D. Zerwas, Talk at PASCOS2012, Merida Mexico, June 3-8, 2012. 

\bibitem{Lee:1977yc} 
  B.~W.~Lee, C.~Quigg and H.~B.~Thacker,
  %``The Strength of Weak Interactions at Very High-Energies and the Higgs Boson Mass,''
  Phys.\ Rev.\ Lett.\  {\bf 38}, 883 (1977).
  %%CITATION = PRLTA,38,883;%%
  See also, D.~A. ~Dicus and V.~S.~Mathur, ~Phys. ~Rev. D7, 3111(1973);   
M. ~Veltman, Acta. Phys. Polon. B8, 475 (1977).

\bibitem{Kane:1992kq} 
  G.~L.~Kane, C.~F.~Kolda and J.~D.~Wells,
  %``Calculable upper limit on the mass of the lightest Higgs boson in any perturbatively valid supersymmetric theory,''
  Phys.\ Rev.\ Lett.\  {\bf 70}, 2686 (1993)
  [hep-ph/9210242].
  %%CITATION = HEP-PH/9210242;%%

\bibitem{can}
  A.~H.~Chamseddine, R.~L.~Arnowitt, P.~Nath,
  %``Locally Supersymmetric Grand Unification,''
  Phys.\ Rev.\ Lett.\  {\bf 49}, 970 (1982);
  %%CITATION = NUB-2559%%
   %``Gauge Hierarchy in Supergravity Guts,''
  Nucl.\ Phys.\  {\bf B227}, 121 (1983);
 %%CITATION = NUPHA,B227,121;%%
  L.~J.~Hall, J.~D.~Lykken, S.~Weinberg,
  %``Supergravity as the Messenger of Supersymmetry Breaking,''
  Phys.\ Rev.\  {\bf D27}, 2359-2378 (1983);
  R.~L.~Arnowitt and P.~Nath,
  %``SUSY mass spectrum in SU(5) supergravity grand unification,''
  Phys.\ Rev.\ Lett.\  {\bf 69}, 725 (1992).
  %%CITATION = PRLTA,69,725;%%
  For a review see, P.~Nath,
    %``Twenty years of SUGRA,''
   [hep-ph/0307123].
%%CITATION = hep-ph/0307123%%

\bibitem{Arbey:2012dq} 
  A.~Arbey, M.~Battaglia, A.~Djouadi and F.~Mahmoudi,
  %``The Higgs sector of the phenomenological MSSM in the light of the Higgs boson discovery,''
  arXiv:1207.1348 [hep-ph].
  %%CITATION = ARXIV:1207.1348;%%

\bibitem{review}
 A.~Djouadi,
  %``The Anatomy of electro-weak symmetry breaking. II. The Higgs bosons in the minimal supersymmetric model,''
  Phys.\ Rept.\  {\bf 459}, 1 (2008);
%  [hep-ph/0503173].
  %%CITATION = HEP-PH/0503173;%%
  M.~S.~Carena and H.~E.~Haber,
  %``Higgs boson theory and phenomenology,''
  Prog.\ Part.\ Nucl.\ Phys.\  {\bf 50}, 63 (2003)

\bibitem{Chan:1997bi}
  K.~L.~Chan \etal%, U.~Chattopadhyay and P.~Nath,
  Phys.\ Rev.\  D {\bf 58} (1998) 096004; 
  %[arXiv:hep-ph/9710473].
  U.~Chattopadhyay, A.~Corsetti and P.~Nath,
  %``WMAP Constraints, SUSY Dark Matter and Implications for the Direct
  %Detection of SUSY,''
  Phys.\ Rev.\  D {\bf 68}, 035005 (2003).
 % [arXiv:hep-ph/0303201].
H.~Baer,  C.~Balazs, A.~Belyaev, T.~Krupovnickas and X.~Tata,
  %``Updated reach of the CERN LHC and constraints from relic density, $b \to s
  %\gamma$ and a($\mu$) in the mSUGRA model,''
  JHEP {\bf 0306}, 054 (2003).
%  [arXiv:hep-ph/0304303].
  %%CITATION = JHEPA,0306,054;%%

\bibitem{Feng:1999mn}
  J.~L.~Feng,  K.~T.~Matchev and T.~Moroi,
  %``Multi - TeV scalars are natural in minimal supergravity,''
  Phys.\ Rev.\ Lett.\  {\bf 84}, 2322 (2000).
%  [arXiv:hep-ph/9908309].
  %%CITATION = PRLTA,84,2322;%%

\bibitem{Feldman:2011ud}
  D.~Feldman, G.~Kane, E.~Kuflik and R.~Lu,
  %``A new (string motivated) approach to the little hierarchy problem,''
  arXiv:1105.3765 [hep-ph].
  %%CITATION = ARXIV:1105.3765;%%

\bibitem{Akula:2011jx} 
  S.~Akula, M.~Liu, P.~Nath and G.~Peim,
  %``Naturalness, Supersymmetry and Implications for LHC and Dark Matter,''
  Phys.\ Lett.\ B {\bf 709}, 192 (2012)
  [arXiv:1111.4589 [hep-ph]].
  %%CITATION = ARXIV:1111.4589;%%

\bibitem{Feldman:2007zn} 
  D.~Feldman, Z.~Liu and P.~Nath,
  %``The Landscape of Sparticle Mass Hierarchies and Their Signature Space at the LHC,''
  Phys.\ Rev.\ Lett.\  {\bf 99}, 251802 (2007);
%  [arXiv:0707.1873 [hep-ph]].
  %%CITATION = ARXIV:0707.1873;%%  
   %``Light Higgses at the Tevatron and at the LHC and Observable Dark Matter in SUGRA and D Branes,''
  Phys.\ Lett.\ B {\bf 662}, 190 (2008);
%  [arXiv:0711.4591 [hep-ph]];
  %%CITATION = ARXIV:0711.4591;%%
%``Sparticles at the LHC,''
  JHEP {\bf 0804}, 054 (2008);
 % [arXiv:0802.4085 [hep-ph]];
  %%CITATION = ARXIV:0802.4085;%%
   Phys.\ Rev.\ D {\bf 78}, 083523 (2008);
  %[arXiv:0808.1595 [hep-ph]].
  %%CITATION = ARXIV:0808.1595;%%
   %``Gluino NLSP, Dark Matter via Gluino Coannihilation, and LHC Signatures,''
  Phys.\ Rev.\ D {\bf 80}, 015007 (2009);
  %[arXiv:0905.1148 [hep-ph]].
  %%CITATION = ARXIV:0905.1148;%%
   %``Low Mass Neutralino Dark Matter in the MSSM with Constraints from $B_s\to \mu^+\mu^-$ and Higgs Search Limits,''
  Phys.\ Rev.\ D {\bf 81}, 117701 (2010).
  %[arXiv:1003.0437 [hep-ph]].
  %%CITATION = ARXIV:1003.0437;%%

\bibitem{Nath:2010zj} 
  P.~Nath, B.~D.~Nelson, H.~Davoudiasl, B.~Dutta, D.~Feldman, Z.~Liu, T.~Han and P.~Langacker {\it et al.},
  %``The Hunt for New Physics at the Large Hadron Collider,''
  Nucl.\ Phys.\ Proc.\ Suppl.\  {\bf 200-202}, 185 (2010)
  [arXiv:1001.2693 [hep-ph]].
  %%CITATION = ARXIV:1001.2693;%%

\bibitem{lowgluino}
 D.~Feldman, K.~Freese, P.~Nath, B.~D.~Nelson and G.~Peim,
  %``Predictive Signatures of Supersymmetry: Measuring the Dark Matter Mass and Gluino Mass with Early LHC data,''
  Phys.\ Rev.\ D {\bf 84}, 015007 (2011);
%  [arXiv:1102.2548 [hep-ph]].
  %%CITATION = ARXIV:1102.2548;%%
 N.~Chen, D.~Feldman, Z.~Liu, P.~Nath and G.~Peim,
  %``Low Mass Gluino within the Sparticle Landscape, Implications for Dark Matter, and Early Discovery Prospects at LHC-7,''
  Phys.\ Rev.\ D {\bf 83}, 035005 (2011);
 % [arXiv:1011.1246 [hep-ph]].
  %%CITATION = ARXIV:1011.1246;%%
   D.~Feldman, Z.~Liu and P.~Nath,
  %``Gluino NLSP, Dark Matter via Gluino Coannihilation, and LHC Signatures,''
  Phys.\ Rev.\ D {\bf 80}, 015007 (2009).
  %[arXiv:0905.1148 [hep-ph]].
  %%CITATION = ARXIV:0905.1148;%%

  \bibitem{lowgluino2}
   M.~A.~Ajaib, T.~Li and Q.~Shafi,
  %``LHC Constraints on NLSP Gluino and Dark Matter Neutralino in Yukawa Unified Models,''
  Phys.\ Lett.\ B {\bf 705}, 87 (2011)
  [arXiv:1107.2573 [hep-ph]].
  %%CITATION = ARXIV:1107.2573;%%  

\bibitem{Baer:2012vr} 
  H.~Baer, V.~Barger, A.~Lessa and X.~Tata,
  %``Discovery potential for SUSY at a high luminosity upgrade of LHC14,''
  arXiv:1207.4846 [hep-ph].
  %%CITATION = ARXIV:1207.4846;%%

\bibitem{georgi}
H. Georgi, in Particles and Fields (edited by C.E. Carlson), A.I.P.,
1975; H. Fritzch and P. Minkowski, Ann. Phys. {\bf 93}, 193 (1975).

\bibitem{Nath:2006ut}
  P.~Nath and P.~Fileviez Perez,
  %``Proton stability in grand unified theories, in strings and in branes,''
  Phys.\ Rept.\  {\bf 441}, 191 (2007).
%  [hep-ph/0601023].
  %%CITATION = HEP-PH/0601023;%%

\bibitem{Babu:2005gx}
  K.~S.~Babu, I.~Gogoladze, P.~Nath and R.~M.~Syed,
  %``A unified framework for symmetry breaking in SO(10),''
  Phys.\ Rev.\  D {\bf 72}, 095011 (2005)
  [arXiv:hep-ph/0506312];
  %%CITATION = PHRVA,D72,095011;%%
%``Fermion mass generation in SO(10) with a unified Higgs sector,''
  Phys.\ Rev.\  D {\bf 74}, 075004 (2006)
  [arXiv:hep-ph/0607244].
  %%CITATION = PHRVA,D74,075004;%%

\bibitem{Nath:2005bx}
  P.~Nath and {R.~M.~Syed},
  %``Couplings of vector-spinor representation for $SO(10)$ model building,''
  JHEP {\bf 0602}, 022 (2006)
  [arXiv: hep-ph/0511172];
  %%CITATION = HEP-PH 0511172;%%
  %``Yukawa Couplings and Quark and Lepton Masses in an SO(10) Model with a
  %Unified Higgs Sector,''
  Phys.\ Rev.\  D {\bf 81}, 037701 (2010)
  [arXiv:0909.2380 [hep-ph]].
  %%CITATION = PHRVA,D81,037701;%%

\bibitem{SU(5)Missingpartner}
 % H.~Georgi,
 % %``An Almost Realistic Gauge Hierarchy,''
 % Phys.\ Lett.\  B {\bf 108}, 283 (1982);
  %%CITATION = PHLTA,B108,283;%%
  A.~Masiero, D.~V.~Nanopoulos, K.~Tamvakis and T.~Yanagida,
  %``Naturally Massless Higgs Doublets In Supersymmetric SU(5),''
  Phys.\ Lett.\  B {\bf 115}, 380 (1982);
  %%CITATION = PHLTA,B115,380;%%
  B.~Grinstein,
  %``A Supersymmetric SU(5) Gauge Theory With No Gauge Hierarchy Problem,''
  Nucl.\ Phys.\  B {\bf 206}, 387 (1982).
  %%CITATION = NUPHA,B206,387;%%

\bibitem{SO(10)Missingpartner}
  K.~S.~Babu, I.~Gogoladze and Z.~Tavartkiladze,
  %``Missing partner mechanism in SO(10) grand unification,''
  Phys.\ Lett.\  B {\bf 650}, 49 (2007)
  [arXiv:hep-ph/0612315].
  %%CITATION = PHLTA,B650,49;%%

\bibitem{Babu:2011tw} 
  K.~S.~Babu, I.~Gogoladze, P.~Nath and R.~M.~Syed,
  %``Variety of SO(10) GUTs with Natural Doublet-Triplet Splitting via the Missing Partner Mechanism,''
  Phys.\ Rev.\ D {\bf 85}, 075002 (2012).
%  [arXiv:1112.5387 [hep-ph]].
  %%CITATION = ARXIV:1112.5387;%%

\bibitem{ms}
R.N. Mohapatra and B. Sakita, Phys. Rev. {\bf D21}, 1062 (1980).

\bibitem{ns}
 P.~Nath and {R.~M.~Syed},
 %``Analysis of couplings with large tensor representations in SO(2N) and  proton decay,''
Phys.\ Lett.\ B {\bf 506}, 68 (2001);
% [arXiv: hep-ph/0103165];
%%CITATION = HEP-PH 0103165;%%
%``Complete cubic and quartic couplings of 16 and 16-bar in SO(10) unification,''
Nucl.\ Phys.\ B {\bf 618}, 138 (2001);
% [arXiv: hep-th/0109116]; ``Coupling the supersymmetric 210 vector multiplet to matter in SO(10),''
Nucl.\ Phys.\ B {\bf 676}, 64 (2004);
% [arXiv: hep-th/0310178];
%%CITATION = HEP-TH 0310178;%%
 R. M. Syed,
 %in {\it Themes in Unification: Pran Nath Festschrift}, ed. by G. Alverson and M.T. Vaughn, (World Scientific, Singapore).
 arXiv: hep-ph/0411054;
 %%CITATION = HEP-PH 0411054;%%
  %``Couplings in SO(10) Grand Unification,''  Ph.D. Thesis, Northeastern University.
   arXiv: hep-ph/0508153.
  %%CITATION = HEP-PH 0508153;%%

 \bibitem{anz}
  P.~Nath and R.~L.~Arnowitt,
  %``Generalized Supergauge Symmetry as a New Framework for Unified Gauge
  %Theories,''
  Phys.\ Lett.\  B {\bf 56}, 177 (1975);
  %%CITATION = PHLTA,B56,177;%%
 R.~L.~Arnowitt, P.~Nath and B.~Zumino,
  %``Superfield Densities and Action Principle in Curved Superspace,''
  Phys.\ Lett.\  B {\bf 56}, 81 (1975).
  %%CITATION = PHLTA,B56,81;%%

\bibitem{fnf}
 D.~Z.~Freedman, P.~van Nieuwenhuizen and S.~Ferrara,
  %``Progress Toward a Theory of Supergravity,''
  Phys.\ Rev.\ D {\bf 13}, 3214 (1976).
  %%CITATION = PHRVA,D13,3214;%%


\bibitem{IbanezRoss} 
  L.~E.~Ibanez and G.~G.~Ross,
  %``Supersymmetric Higgs and radiative electroweak breaking,''
  Comptes Rendus Physique\ {\bf 8}, 1013  (2007).
  %[hep-ph/0702046 [HEP-PH]].
  %%CITATION = CRPOB,8,1013;%%

\bibitem{Martin:1993zk} 
  S.~P.~Martin and M.~T.~Vaughn,
  %``Two loop renormalization group equations for soft supersymmetry breaking couplings,''
  Phys.\ Rev.\ D {\bf 50}, 2282 (1994)
%  [Erratum-ibid.\ D {\bf 78}, 039903 (2008)]
%  [hep-ph/9311340].
  %%CITATION = HEP-PH/9311340;%%

\bibitem{Kors:2003wf} 
  B.~Kors and P.~Nath,
  %``Effective action and soft supersymmetry breaking for intersecting D-brane models,''
  Nucl.\ Phys.\ B {\bf 681}, 77 (2004);
 % [hep-th/0309167].
  %%CITATION = HEP-TH/0309167;%%
 D.~Lust, S.~Reffert and S.~Stieberger,
  %``Flux-induced soft supersymmetry breaking in chiral type IIB orientifolds with D3 / D7-branes,''
  Nucl.\ Phys.\ B {\bf 706}, 3 (2005).
%  [hep-th/0406092].
  %%CITATION = HEP-TH/0406092;%%

\bibitem{Chattopadhyay:2001mj}
  U.~Chattopadhyay and P.~Nath,
  %``b - tau unification, g(mu) - 2, the b ---> s + gamma constraint and nonuniversalities,''
  Phys.\ Rev.\ D {\bf 65}, 075009 (2002).
%  [hep-ph/0110341].
  %%CITATION = HEP-PH/0110341;%%

  \bibitem{DPRoy}
  D. P.Roy, Talk at PASCOS2012, Merida Mexico, June 3-8, 2012  

\bibitem{Nath:1997qm}
  P.~Nath, R.~L.~Arnowitt,
  %``Nonuniversal soft SUSY breaking and dark matter,''
  Phys.\ Rev.\  {\bf D56}, 2820-2832 (1997);
%  [hep-ph/9701301].
  %%CITATION = HEP-PH/9701301;%%  
  J.~R.~Ellis, K.~A.~Olive, Y.~Santoso,
  %``The MSSM parameter space with nonuniversal Higgs masses,''
  Phys.\ Lett.\  {\bf B539}, 107-118 (2002).
 % [hep-ph/0204192].

\bibitem{Akula:2011dd} 
  S.~Akula, D.~Feldman, Z.~Liu, P.~Nath and G.~Peim,
  %``New Constraints on Dark Matter from CMS and ATLAS Data,''
  Mod.\ Phys.\ Lett.\ A {\bf 26}, 1521 (2011)
  [arXiv:1103.5061 [hep-ph]].
  %%CITATION = ARXIV:1103.5061;%%

\bibitem{wmap}
  E.~Komatsu {\it et al.}  [WMAP Collaboration],
  %``Seven-Year Wilkinson Microwave Anisotropy Probe (WMAP) Observations: Cosmological Interpretation,''
  Astrophys.\ J.\ Suppl.\  {\bf 192}, 18 (2011).
%  [arXiv:1001.4538 [astro-ph.CO]].
  %%CITATION = ARXIV:1001.4538;%%

\bibitem{Feldman:2010wy}
  D.~Feldman, Z.~Liu, P.~Nath and G.~Peim,
  %``Multicomponent Dark Matter in Supersymmetric Hidden Sector Extensions,''
  Phys.\ Rev.\  D {\bf 81}, 095017 (2010).
%  [arXiv:1004.0649 [hep-ph]].
  %%CITATION = PHRVA,D81,095017;%%

\bibitem{Akula:2012kk} 
  S.~Akula, P.~Nath and G.~Peim,
  %``Implications of the Higgs Boson Discovery for mSUGRA,''
  arXiv:1207.1839 [hep-ph].
  %%CITATION = ARXIV:1207.1839;%%

\bibitem{xenon}
  E.~Aprile {\it et al.}  [XENON100 Collaboration],
  %``Dark Matter Results from 100 Live Days of XENON100 Data,''
  arXiv:1104.2549 [astro-ph.CO];
  %%CITATION = ARXIV:1104.2549;%%
Phys.\ Rev.\ Lett.\  {\bf 105}, 131302 (2010);
   %``Likelihood Approach to the First Dark Matter Results from XENON100,''
  arXiv:1103.0303 [hep-ex].
  %%CITATION = ARXIV:1103.0303;%%

  \bibitem{futureXENON}
E. Aprile, The XENON Dark Matter Search, WONDER Workshop, LNGS, March 22, 2010.

\bibitem{futureSCDMS}
B.~Cabrera, ``SuperCDMS Development Project",  2005.

\bibitem{Yuan:1984ww} 
  T.~C.~Yuan, R.~L.~Arnowitt, A.~H.~Chamseddine and P.~Nath,
  %``Supersymmetric Electroweak Effects on G-2 (mu),''
  Z.\ Phys.\ C {\bf 26}, 407 (1984);
  %%CITATION = ZEPYA,C26,407;%%
D.~A.~Kosower \etal%, L.~M.~Krauss and N.~Sakai,
  %``Low-Energy Supergravity and the Anomalous Magnetic Moment of the Muon,''
  Phys.\ Lett.\ B {\bf 133}, 305 (1983).
  %%CITATION = PHLTA,B133,305;%%

\bibitem{hoecker}
A.~Hoecker,
  %``The Hadronic Contribution to the Muon Anomalous Magnetic Moment and to the Running Electromagnetic Fine Structure Constant at MZ - Overview and Latest Results,''
  Nucl.\ Phys.\ Proc.\ Suppl.\  {\bf 218}, 189 (2011);
%  [arXiv:1012.0055 [hep-ph]].
  %%CITATION = ARXIV:1012.0055;%%
    K.~Hagiwara \etal%, R.~Liao, A.~D.~Martin, D.~Nomura and T.~Teubner,
  %``(g-2)_mu and alpha(M_Z^2) re-evaluated using new precise data,''
  J.\ Phys.\ G G {\bf 38}, 085003 (2011).
%  [arXiv:1105.3149 [hep-ph]].
  %%CITATION = ARXIV:1105.3149;%%
%
%%

\bibitem{Babu:2008ge} 
  K.~S.~Babu, I.~Gogoladze, M.~U.~Rehman and Q.~Shafi,
  %``Higgs Boson Mass, Sparticle Spectrum and Little Hierarchy Problem in Extended MSSM,''
  Phys.\ Rev.\ D {\bf 78}, 055017 (2008)
  [arXiv:0807.3055 [hep-ph]].
  %%CITATION = ARXIV:0807.3055;%%

\bibitem{adm1}
S. Nussinov, Phys. Lett. B {\bf 165}, 55 (1985);
K. Griest and D. Seckel.
%Cosmic Asymmetry, Neutrinos and the Sun.
Nucl. Phys. B {\bf 283}, 681 (1987);
R.S. Chivukula and T.P. Walker, Nucl. Phys. B {\bf 329}, 445 (1990);
  S.~Dodelson, B.~R.~Greene and L.~M.~Widrow,
  %``Baryogenesis, dark matter and the width of the Z,''
  Nucl.\ Phys.\ B {\bf 372}, 467 (1992);
  %%CITATION = NUPHA,B372,467;%%
  S.~M.~Barr,
  %``Baryogenesis, sphalerons and the cogeneration of dark matter,''
  Phys.\ Rev.\ D {\bf 44}, 3062 (1991);
  %%CITATION = PHRVA,D44,3062;%%
D.~B.~Kaplan,
  %``A Single explanation for both the baryon and dark matter densities,''
  Phys.\ Rev.\ Lett.\  {\bf 68}, 741 (1992).
  %%CITATION = PRLTA,68,741;%%

\bibitem{Davoudiasl:2012uw}
H.~Davoudiasl and R.~N.~Mohapatra,
%``On Relating the Genesis of Cosmic Baryons and Dark Matter,''
arXiv:1203.1247 [hep-ph].
%%CITATION = ARXIV:1203.1247;%%

\bibitem{admBL}
  D.~E.~Kaplan, M.~A.~Luty and K.~M.~Zurek,
  %``Asymmetric Dark Matter,''
  Phys.\ Rev.\ D {\bf 79}, 115016 (2009).
%  [arXiv:0901.4117 [hep-ph]].
  %%CITATION = ARXIV:0901.4117;%%

\bibitem{ht}
J.~A.~Harvey and M.~S.~Turner
  %``Cosmological baryon and lepton number in the presence of electroweak fermion-number violation,''
  Phys.\ Rev.\ D {\bf 42}, 3344 (1990).

  \bibitem{fnp}
    W.~-Z.~Feng, P.~Nath and G.~Peim,
  %``Cosmic Coincidence and Asymmetric Dark Matter in a Stueckelberg Extension,''
  Phys.\ Rev.\ D {\bf 85}, 115016 (2012).
%  [arXiv:1204.5752 [hep-ph]].
  %%CITATION = ARXIV:1204.5752;%%

\bibitem{Martin:1992mq} 
  S.~P.~Martin,
  %``Some simple criteria for gauged R-parity,''
  Phys.\ Rev.\ D {\bf 46}, 2769 (1992).
%  [hep-ph/9207218].
  %%CITATION = HEP-PH/9207218;%%

\bibitem{Barger:2008wn} 
  V.~Barger, P.~Fileviez Perez and S.~Spinner,
  %``Minimal gauged U(1)(B-L) model with spontaneous R-parity violation,''
  Phys.\ Rev.\ Lett.\  {\bf 102}, 181802 (2009).
%  [arXiv:0812.3661 [hep-ph]].
  %%CITATION = ARXIV:0812.3661;%%

\bibitem{Perez}
P. Fileviez Perez, Talk at PASCOS2012, Merida Mexico, June 3-8, 2012. 

\bibitem{kn12}
B.~K\"ors and P.~Nath,
%{\it A Stueckelberg Extension of the Standard Model},
Phys.\ Lett.\ B {\bf 586} (2004) 366;
%[hep-ph/0402047];
%%CITATION = HEP-PH 0402047;%%
%B.~K\"ors and P.~Nath,
%{\it A Supersymmetric Stueckelberg U(1) Extension of the MSSM},
JHEP {\bf 0412} (2004) 005;
 %[hep-ph/0406167];
%%CITATION = HEP-PH 0406167;%%
  JHEP {\bf 0507}, 069 (2005).
  %[arXiv:hep-ph/0503208].
  %%CITATION = JHEPA,0507,069;%

\bibitem{FLNPRL}
  D.~Feldman, Z.~Liu and P.~Nath,
  %``Probing a Very Narrow $Z'$ Boson with CDF and D0 Data,''
  Phys.\ Rev.\ Lett.\  {\bf 97}, 021801 (2006);
  %[arXiv:hep-ph/0603039];
  %%CITATION = PRLTA,97,021801;%%
 D.~Feldman, B.~Kors and P.~Nath,
  %``Extra-weakly Interacting Dark Matter,''
  Phys.\ Rev.\  D {\bf 75}, 023503 (2007);
%[arXiv:hep-ph/0610133].
  %%CITATION = PHRVA,D75,023503;%%
D.~Feldman, Z.~Liu and P.~Nath,
%{\it The Stueckelberg Z' extension with kinetic mixing and milli-charged dark
  %matter from the hidden sector},
  Phys.\ Rev.\  D {\bf 75}, 115001 (2007);
%  [arXiv:hep-ph/0702123].
  %%CITATION = PHRVA,D75,115001;%%
  D.~Feldman, P.~Fileviez Perez and P.~Nath,
  %``R-parity Conservation via the Stueckelberg Mechanism: LHC and Dark Matter Signals,''
  JHEP {\bf 1201}, 038 (2012).
%  [arXiv:1109.2901 [hep-ph]].
  %%CITATION = ARXIV:1109.2901;%%

  \bibitem{Ringwald}
  A. Ringwald, Talk at PASCOS2012, Merida Mexico, June 3-8, 2012. 

\bibitem{Hewett:2012ns} 
  J.~L.~Hewett, H.~Weerts, R.~Brock, J.~N.~Butler, B.~C.~K.~Casey, J.~Collar, A.~de Govea and R.~Essig {\it et al.},
  %``Fundamental Physics at the Intensity Frontier,''
  arXiv:1205.2671 [hep-ex].
  %%CITATION = ARXIV:1205.2671;%%

\bibitem{Abe:2011ts} 
  K.~Abe, T.~Abe, H.~Aihara, Y.~Fukuda, Y.~Hayato, K.~Huang, A.~K.~Ichikawa and M.~Ikeda {\it et al.},
  ``Letter of Intent: The Hyper-Kamiokande Experiment --- Detector Design and Physics Potential ---,''
  arXiv:1109.3262 [hep-ex].
  %%CITATION = ARXIV:1109.3262;%%

\bibitem{Ibrahim:2007fb} 
  T.~Ibrahim and P.~Nath,
  %``CP Violation From Standard Model to Strings,''
  Rev.\ Mod.\ Phys.\  {\bf 80}, 577 (2008).
%  [arXiv:0705.2008 [hep-ph]].
  %%CITATION = ARXIV:0705.2008;%%

\bibitem{Arnowitt:1993pd} 
  R.~L.~Arnowitt and P.~Nath,
  %``Testing Supergravity Grand Unification at Future Accelerator and Underground Experiments,''
  Phys.\ Rev.\ D {\bf 49}, 1479 (1994).
%  [hep-ph/9309252].
  %%CITATION = HEP-PH/9309252;%%

\end{thebibliography}
  \end{document}